\DocumentMetadata{
	lang		= eng-US, 	
	pdfstandard	= A-3u,		
	pdfversion	= 1.7, 		
}

\documentclass[twocolumn]{article}
\usepackage[margin=1in]{geometry}
\usepackage{hyperref}
\usepackage{amsmath,amssymb}
\usepackage{graphicx}
\usepackage{subcaption}
\usepackage{newtxtext,newtxmath} 

\usepackage{algorithm}
\usepackage{algorithmic}
\usepackage{booktabs}
\usepackage{xcolor}
\usepackage{colortbl}
\usepackage{array}
\usepackage{float}
\usepackage{multirow}
\usepackage{enumitem}
\usepackage[numbers,sort&compress]{natbib}
\makeatletter

\@addtoreset{paragraph}{subsubsection}

\renewcommand{\paragraph}{%
  \@startsection{paragraph}{4}{\z@}%
  {3.25ex \@plus1ex \@minus.2ex}
  {-1em}
  {\normalfont\normalsize\bfseries}%
}

\makeatother

\allowdisplaybreaks 

\begin{document}

\title{Bridging CAD and Data-Driven Design: Attributed Feature 
Graphs for Engineering Design\thanks{Preprint.}}

\author{
  Abhishek Indupally$^{1}$\thanks{Corresponding author: aindupa@clemson.edu} \and
  Ibraheem Alawadhi$^{2}$ \and
  Satchit Ramnath$^{1}$ \and
  Jami J. Shah$^{2}$
}

\date{
  $^{1}$Clemson University, Clemson, SC \\
  $^{2}$The Ohio State University, Columbus, OH
}

\maketitle

\maketitle



\begin{abstract} \label{abstract}
Engineering design is an iterative, simulation‑driven process where traditional workflows rely heavily on computationally expensive analyses such as finite element and computational fluid dynamics. Although data‑driven methods have accelerated design evaluation and optimization, most existing geometric representations discard parametric and feature‑level semantics, limiting their integration with CAD‑driven design workflows and reducing model interpretability. To address this gap, this work introduces Attributed Feature Graphs (AFGs), a feature‑based representation that encodes design features, such as extrusions, ribs, and pockets, as nodes and their geometric or dependency relations as directed edges. AFGs preserve design intent and parametric structure while remaining compatible with standard graph‑based learning methods, enabling end‑to‑end learning directly on CAD‑derived feature graphs. The paper demonstrates the proposed representation through a surrogate‑modeling case study on the CarHoods10K automotive hood frame dataset, where a Graph Neural Network (GNN) is trained as an evaluation engine to predict performance metrics from AFG inputs. The learned model achieves competitive surrogate performance compared with traditional data‑driven approaches, but with the added benefit that engineers can map predictions back to specific CAD features and interpret how individual design elements influence system behavior. Furthermore, because AFGs are built from native CAD features, engineers can directly edit the underlying geometry in the CAD environment and re‑evaluate the design through the same learned model.

\end{abstract}

            \begin{figure*}
                \centering\includegraphics[width=\textwidth, alt = {Different Types of 3D Representations }]{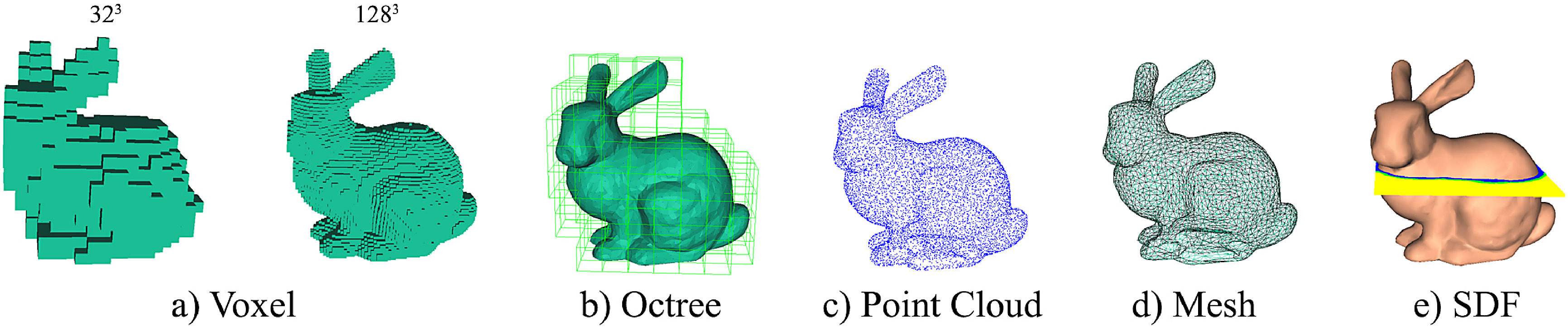}
            \caption{Different Types of 3D Representations \cite{fahim_single-view_2021} }\label{fig:3drepresentations}
            \end{figure*}
             
\section{Introduction}
\label{sec:Intro}
    Engineering design is a complex, iterative process where traditional workflows rely heavily on simulation-driven loops such as finite element analysis (FEA) or computational fluid dynamics (CFD) that are computationally expensive and time-consuming, particularly during conceptual and preliminary stages. Data-driven methods have emerged as powerful alternatives, accelerating diverse tasks across design stages, i.e., generation, evaluation, and optimization by using Artificial Intelligence (AI) / Machine Learning (ML) to rapidly process early concepts and inform targeted iterations \cite{khanolkar_mapping_2023}. 
    
    As the use and research of these methods grow exponentially, the way data is represented to these models becomes increasingly important.  In fields such as computer vision, 3D modeling in civil engineering, and animation, researchers have extensively explored geometric representations such as meshes, point clouds, and triplanes. However, these representations often lack the parametric information essential for engineering design applications. Engineering design is inherently iterative, with engineers continuously exploring concepts, testing performance across use cases, and refining designs based on results. To support this process, representations must enable easy modification and reusability of design elements. Feature-based design (FBD) addresses this need \cite{camba_parametric_2016}, allowing engineers to capture, edit, and manipulate geometric features parametrically (e.g.,  feature depth, fillet radius, offset values), ensuring both flexibility and traceability throughout the design cycle. Because FBD captures design intent and supports higher‑level design decisions (e.g., modifying feature types, dimensions, or sequences), it provides a natural foundation for training AI/ML models that operate directly on geometric features. 
    
    Recent advances explore diverse representations for training various ML methods such as Convolution Neural Networks (CNNs), Variational Auto Encoders (VAEs), and Graph Neural Networks (GNNs)~\cite{kumar_study_2024, krahe_ai-based_2021, indupally_developing_2025, ahmed_graph_2021, mahajan_orthocad-322k_2025, lai_graphbrep_2026, li_brepgpt_2025, wang_normalnet_2019, maturana_voxnet_2015}, for a range of tasks including design generation, optimization, and evaluation, prioritizing accuracy and intra‑domain generalization. However, these representations often retain little or no explicit parametric or feature‑level semantics, limiting their direct integration with CAD‑driven design workflows. Geometric representations such as B‑Rep, meshes, voxels, and point clouds enable ML training but discard CAD‑native semantics such as feature intent, parametric relations, and manufacturability, thereby hindering interpretability and the critical CAD‑AI feedback loop during iterative design.
    
    This paper introduces \emph{Attributed Feature Graphs} (AFGs), which extract CAD features as scalable graphs (nodes: features with dimensional attributes; edges: topological and positional relations) to preserve engineering intent for AI/ML tasks. This paper demonstrates AFGs on an automotive hood frame dataset (CarHoods10K)~\cite{wollstadt_carhoods10k_2022, noauthor_dryad_nodate}, where AFGs trained on GNNs achieve performance comparable to baseline representations in a performance-prediction task while enabling seamless re‑entry into CAD for editing and refinement. This work helps close gaps in data representation, enabling faster progress through data‑driven decisions while preserving engineers’ freedom to shape and refine designs throughout the iterative process.

	Section~\ref{background} reviews different 3D representations used in engineering design and  summarizes how AI/ML is applied across design stages. Section~\ref{methodology-afg} details the Attributed Feature Graph (AFG) methodology. Section~\ref{case-study} presents a surrogate‑modeling case study using AFGs, and Section~\ref{discussion} discusses advantages, limitations, and future work.


\section{Background}\label{background}
    \subsection{3D Representations in ML}
    \label{3drep}
    Three-dimensional geometry representations underpin computational design, FEA/CFD, and generative modeling in mechanical engineering. Explicit representations (B-reps, meshes, point clouds, voxels) provide CAD interoperability and simulation readiness but suffer from topological rigidity. Implicit representations (SDFs, neural fields) enable differentiable optimization yet challenge CAD extraction. Hybrid approaches bridge these paradigms. This review synthesizes ML applications across representations, tracing evolution from classification to generative/surrogate modeling. An overview of the key representations used is summarized in Table \ref{tab:summary-3d-representations}.
          \subsubsection{Explicit Representations}
          \label{3drep:explicit}
            In mechanical engineering, explicit 3D geometry representations serve as the foundational encoding for design, analysis, and optimization workflows, ranging from parametric CAD modeling to finite element simulations (FEA) and computational fluid dynamics (CFD). These representations directly encode surfaces, volumes, or elements through discrete or combinatorial structures, making them the dominant choice in industrial CAD/CAE tools for their precision and editability. The following subsections first describe the geometric structure of key explicit representations and then summarize how researchers have employed them in machine learning tasks such as classification, generation, optimization, and surrogate modeling.
            \paragraph{Boundary Representations (B-Reps)} 
            Boundary representations (B‑Reps) encode solids through parametric surfaces (e.g., NURBS, B‑splines) organized into watertight faces, edges, and vertices with explicit topological adjacency. B‑Reps dominate mechanical CAD systems because they support precise downstream analysis (FEA, CFD) and manufacturing, yet their combinatorial complexity and topological constraints make them challenging for direct learning.

            Recent work leverages the native B‑Rep structure for generative modeling and analysis. UV‑Net \cite{jayaraman_uv-net_2021} constructs face‑adjacency graphs and samples parametric surfaces via UV grids, enabling 2D convolutional operations on CAD‑native geometry for classification, segmentation, and retrieval. SolidGen \cite{jayaraman_solidgen_2022} treats serialized B‑Rep hierarchies (vertex–edge–face arrays) as sequences and employs autoregressive transformers to generate valid solids, supporting topologically complex designs. BrepGen \cite{xu_brepgen_2024} embeds parametric features and mating constraints in hierarchical latent trees, which are denoised top‑down to produce valid B‑Reps while preserving geometric constraints. GraphBRep \cite{lai_graphbrep_2026} decouples topology (via adjacency matrices) from geometry and uses a staged diffusion model to generate discrete topology followed by continuous surface parameters, improving validity and efficiency. BrepGPT \cite{li_brepgpt_2025} tokenizes half‑edge neighborhoods via Voronoi half‑patches to convert B‑Reps into discrete sequences, enabling transformer‑based end‑to‑end CAD solid generation through 1D sequence modeling. Despite these advances, these methods operate on topology and geometry rather than explicit CAD feature trees and construction intent, limiting their ability to integrate cleanly with feature‑based design workflows.

            \paragraph{Mesh Representations}
            A mesh discretizes a surface or volume via vertices, edges, and faces, commonly using triangles or quadrilaterals as shown in Figure \ref{fig:3drepresentations}d. In mechanical analysis, models are typically meshed with 3D hexahedral or tetrahedral elements for detailed simulations, or 1D beam/2D shell elements in early‑stage analysis. Meshes record explicit connectivity and geometry, enabling efficient finite‑element and boundary‑element methods but discarding the original CAD parametric history.
            Researchers have used meshes for classification, generation, and surrogate modeling. MeshPointNet \cite{heyrani_nobari_meshpointnet_2024} combines GNN‑like convolutions with PointNet‑style point‑wise processing on aircraft meshes, improving surface classification and supporting uncertainty‑aware CFD meshing. Salimath et al. \cite{salimath_hybrid_2023} transform mesh node data (e.g., forging wear) into graph‑like representations and apply GNNs to surrogate FEM simulations, achieving orders‑of‑magnitude speedups. MeshCNN \cite{hanocka_meshcnn_2019} introduces edge‑centric convolutions on irregular meshes, defining convolution kernels over face‑sharing neighborhoods and enabling permutation‑equivariant shape classification and segmentation. MeshGPT \cite{siddiqui_meshgpt_2024} quantizes mesh triangles into discrete tokens via VQ‑VAE and treats mesh modeling as a token‑sequence generation task, using decoder‑only transformers to generate realistic mesh distributions in a language‑modeling framework.
            These methods demonstrate that meshes can support scalable, physics‑aware surrogates, but because they drop parametric and feature‑level semantics, mapping mesh‑level predictions back to editable CAD operations remains challenging.
        
            \paragraph{Point Cloud Representations} A point cloud is an unstructured set of 3D points, each with spatial coordinates $(x,y,z)$ and optionally surface normals or color as shown in Figure \ref{fig:3drepresentations}c. Point clouds often represent the external surface of scanned objects or scenes and naturally capture imperfections, discontinuities, and sensor‑induced holes or noise. They are lightweight and easy to generate from LiDAR or photogrammetry but lack explicit connectivity or topology, which complicates robust finite‑element simulation and reverse‑engineering.

            In ML, point‑cloud representations have been used for classification, segmentation, and process identification. PointNet and successors treat points as an unordered set and apply shared MLPs and max‑pooling to capture global shape descriptors \cite{charles_pointnet_2017}. MRIConv++ and PRIConv++ classify manufacturing processes from point‑wise features, achieving high accuracy (98.5–99.6\%) while avoiding the computational overhead of voxelization \cite{jaiswal_mesh-driven_nodate, liu_manufacturing_2025, li_deep_2021}. Point‑based methods often outperform low‑resolution voxel representations in dense data tasks and have been extended toward implicit embeddings that recover topology and boundary structure at the cost of increased labeling requirements \cite{li_deep_2021}.
            
            Despite their efficiency and fidelity to raw scans, point clouds lack explicit feature structure and connectivity, which limits their direct use in parametric CAD workflows where design intent and manufacturability constraints must be preserved.

\begin{table*}[htbp]
\centering
\small
\caption{Summary of 3D Representations in Mechanical Engineering Design and ML}
\label{tab:summary-3d-representations}
\resizebox{\textwidth}{!}{%
\begin{tabular}{p{1.8cm}p{2.8cm}p{3.5cm}p{3.8cm}p{3.8cm}}
            \toprule
            \textbf{Type} & \textbf{Core Structure} & \textbf{Strengths} & \textbf{Weaknesses} & \textbf{Key ML/Engineering Applications} \\
            \midrule
            \rowcolor{blue!10}
            \textbf{B‑Rep} & Parametric surfaces (NURBS) + watertight topology & Precision, CAD/CAE compatibility & Combinatorial complexity & UV‑Net~\cite{jayaraman_uv-net_2021}, SolidGen~\cite{jayaraman_solidgen_2022}, BrepGen~\cite{xu_brepgen_2024} \\
            \rowcolor{blue!10}
            \textbf{Mesh} & Vertices/edges/faces (tri/quad) & FEM/CFD ready, explicit connectivity & No parametric history & MeshCNN~\cite{hanocka_meshcnn_2019}, MeshGPT~\cite{siddiqui_meshgpt_2024} \\
            \rowcolor{blue!10}
            \textbf{Point Cloud} & Unstructured 3D points (± normals) & Lightweight, scan‑native & No topology/connectivity & PointNet~\cite{charles_pointnet_2017}, MRIConv++~\cite{jaiswal_mesh-driven_nodate} \\
            \rowcolor{blue!10}
            \textbf{Voxel/Grid} & 3D occupancy/density grids & Simple 3D CNNs & Resolution/memory limits & VoxNet~\cite{maturana_voxnet_2015}, X‑Cube~\cite{ren_xcube_2024} \\
            \rowcolor{blue!10}
            \textbf{Feature‑Based (FBD)} & Engineering features (holes, ribs) + intent & Design rationale, parametric edits & Limited ML datasets &  HPSketch \cite{fan_history-based_2025}, RNNs \cite{krahe_ai-based_2021, krahe_ai-based_2021-1} \\
            \midrule[1.5pt]
            \rowcolor{green!8}
            \textbf{Analytical Implicits} & \(f(\mathbf{x})=0\) (quadrics, metaballs) & Exact eval, stable booleans & Limited primitives, polygonization & nTop/Altair field-based \cite{noauthor_implicit_nodate, noauthor_lets_nodate}, CAD kernels \\
            \rowcolor{green!12}
            \textbf{Neural Implicits (INR)} & MLP: \(f(x)=0\) (SDF/occupancy) & Differentiable, generative & Training cost, sharp edges & DeepSDF~\cite{park_deepsdf_2019}, Karki SBM/CFD~\cite{karki_mechanics_2026} \\
            \midrule[1.5pt]
            \rowcolor{orange!10}
            \textbf{Hybrid} & Explicit + implicit combinations & Complementary strengths & Implementation complexity & Point2CAD~\cite{liu_point2cad_2024}, BR‑DF~\cite{zhang_b-rep_2025}, HybridSDF~\cite{vasu_hybridsdf_2022} \\
            \bottomrule
\end{tabular}}
\end{table*}               
            \paragraph{Voxel/Grid/Vectorized Design Representations} Voxel representations (figure \ref{fig:3drepresentations}a divide 3D space into a regular grid of cubic cells, storing occupancy, material, or density values. They extend 2D pixel‑based convolutions into 3D, enabling straightforward implementation of 3D CNNs for shape analysis and generation, but suffer from cubic memory growth and resolution limits. VoxNet \cite{maturana_voxnet_2015} converts LiDAR point clouds into 3D occupancy grids and applies 3D CNNs for real‑time object recognition, capturing spatial hierarchies from noisy data. NormalNet \cite{wang_normalnet_2019} augments binary occupancy voxels with surface normals, mapping orientation vectors directly onto voxels and using specialized convolution modules to improve 3D object classification and retrieval. X‑Cube \cite{ren_xcube_2024} addresses the memory limits of dense voxel grids by using sparse voxel hierarchies and a cascaded diffusion architecture; a sparse VAE paired with multi‑resolution diffusion models enables large‑scale 3D generation while preserving fine geometric details. Although these methods demonstrate the utility of voxel‑based encodings for classification and generation, they typically discard explicit feature‑level semantics and parametric structure, making it difficult to map learned outputs directly back to editable CAD features and manufacturing‑oriented design choices.
        
\begin{figure}[H]
            \centering\includegraphics[width=\linewidth, alt = {Feature-Based Modeling }]{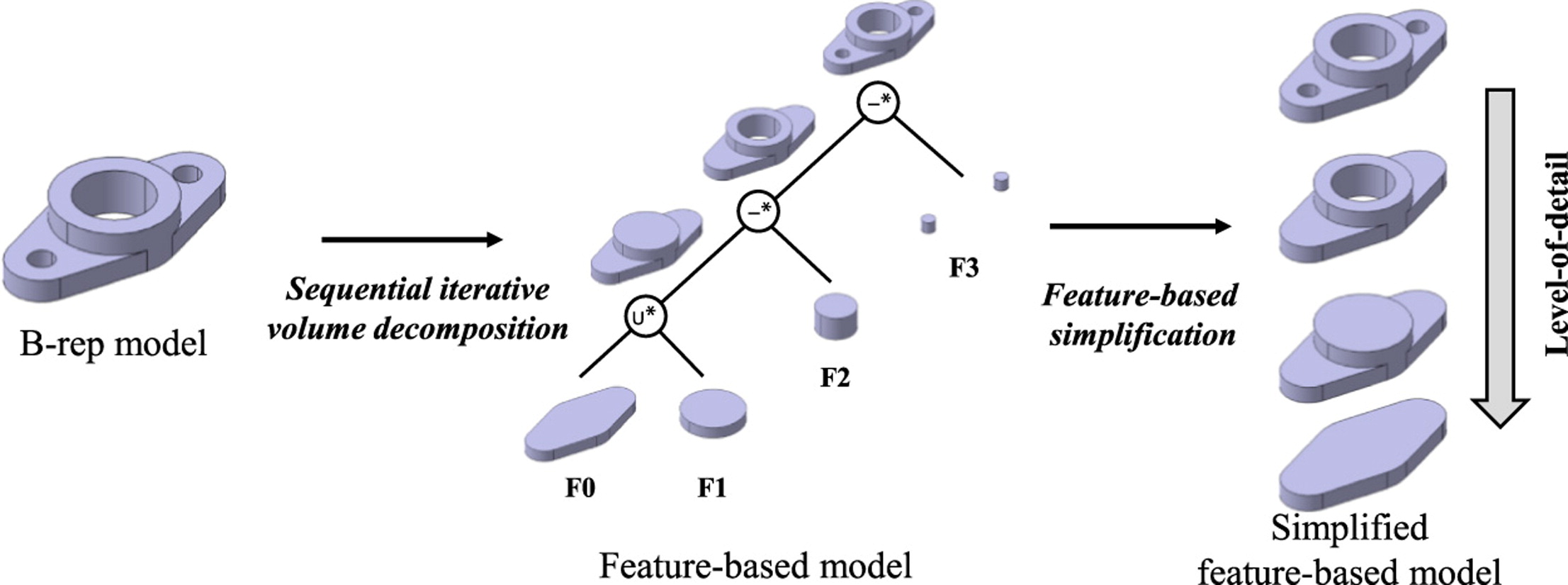}
            \caption{Feature-Based Design Approach \cite{kim_feature-based_2014}}\label{fig:fbd_brep}
            \end{figure}

            \paragraph{Feature-Based Representations} 
            Feature-Based Design (FBD) encodes engineering intent through parameterized, semantically-rich features (holes, pockets, ribs, slots) that capture both geometric form and functional purpose~\cite{salomons_review_nodate, shah_parametric_1995} as shown in Figure~\ref{fig:fbd_brep}. Unlike B-Rep/CSG's final-shape focus, FBD addresses a key limitation of geometric CAD models: the absence of explicit design rationale and process semantics~\cite{wang_scheme_nodate}. Organized as construction sequences with a priori (conceptual) or a posteriori (recognition) methodologies, FBD bridges functional intent and downstream applications but relies on predefined feature parameterizations and admissible operations, which can limit its expressiveness for freeform geometries and highly unconventional forms. In addition, the representation of complex models as ordered feature histories can introduce computational overhead as the number of primitives and construction operations increases, affecting scalability for large 3D shape instances. Although FBD is well suited to parametric, feature-based CAD models, it lacks extensive ML-ready datasets for learning feature hierarchies. The significance of FBD lies in its ability to bridge functional design intent, geometric representation, and downstream engineering applications.            
            
            
        \subsubsection{Implicit Representations}
        \label{3drep:implicit}
        Beyond explicit geometric encodings such as B‑reps, meshes, and point clouds—long‑established in CAD for their editability but limited by resolution and topological rigidity—implicit representations provide continuous, differentiable alternatives that enhance topology flexibility and compatibility with ML‑driven design workflows. These representations model 3D geometry as the zero-level set of scalar fields \(f(x) = 0,\) where \(\ x \in R^{3}\ \) denotes spatial coordinates, typically parameterized by neural networks that output signed distances, occupancy probabilities, or other geometric properties. This formulation enables resolution-independent reconstructions, smooth topology transitions, and automatic differentiation, making implicit representations particularly valuable for generative modeling, surrogate prediction, and meshless simulation tasks where traditional discretization proves limiting.
                
            \paragraph{Non-Neural Implicits} 
            Non-neural implicits define geometry via analytical scalar fields \(f(\mathbf{x}) = 0\) (quadrics, metaballs, superquadrics), enabling exact primitives and smooth blends through soft-min/max operations without learned networks \cite{bloomenthal_implicit_1994, bloomenthal_introduction_1997}. Industrial field-based systems (nTop/Altair) drive lattices and variable-density structures for AM/topology optimization with robust booleans and offsets \cite{noauthor_implicit_nodate, noauthor_lets_nodate}. These provide simulation-ready evaluation on analysis grids and integrate with engineering workflows via level-set extraction \cite{zhang_neural_2025}. ML applications use analytical fields as surrogates or hybrid GNN inputs for CAD retrieval. Compared to neural INRs, non-neurals guarantee computational stability and parallel evaluation but require polygonization (introducing artifacts) and lack generative flexibility \cite{grimm_simple_2002}.
        
            \paragraph{Neural Implicits} 
            Signed distance functions (SDFs) map spatial points \( x \in \mathbb{R}^3 \) to signed distances from the nearest surface (\(f(x) = 0\)), while implicit neural representations (INRs) parameterize continuous fields via MLPs for dynamic querying. DeepSDF~\cite{park_deepsdf_2019} introduced neural SDFs with auto-decoder MLPs using latent codes for unsupervised reconstruction, surpassing discrete limits. Engineering applications include Lipschitz-regularized INRs for mechanical brackets~\cite{rebbagondla_neural_nodate} and BladeSDF for turbine blades with strain-conditioned generation.
            INRs couple with shifted boundary method (SBM) solvers for meshless mechanics/CFD surrogates~\cite{karki_mechanics_2026,karki_direct_2025} and IGNNS~\cite{xu_implicit_2025} predict mesh size functions via GAN-sampled SDF grids. Eikonal enforcement (\(\|\nabla f\| = 1\)) ensures gradient fidelity. INRs excel in resolution-independent smoothness and end-to-end optimization but struggle with sharp edges, training stability, and CAD extraction.
            
        \subsubsection{Hybrid Representations} Hybrid representations lack a single strict definition but generally combine multiple geometric encodings to exploit complementary strengths—including explicit‑explicit (e.g., meshes + voxels), implicit‑implicit (e.g., SDF + occupancy), and explicit‑implicit mixtures (e.g., point clouds + neural fields). In ML workflows for engineering design, hybrids address the limitations of pure forms by coupling explicit topology with implicit smoothness or volumetric interiors with surface fidelity, facilitating reconstruction, prediction, and generation tasks where purely explicit or purely implicit encodings fall short. IGNN \cite{xu_implicit_2025} samples continuous SDFs onto voxel grids as GAN inputs to regress adaptive mesh size functions. This implicit-implicit hybrid preserves SDF differentiability while enabling structured meshing. Local Deep Implicit Functions (LDIF) \cite{genova_local_2020} decomposes shapes into explicit local partitions modulated by Gaussian-residual neural fields. The approach scales reconstruction for fine details beyond global implicits. Neural Sparse Voxel Fields (NSVF) \cite{liu_neural_2020} bounds neural radiance fields within pruned voxel octrees for novel view synthesis. Explicit voxel structure accelerates rendering by skipping empty space. Point2CAD \cite{liu_point2cad_2024} segments point clouds explicitly then fits INR freeforms for B-rep reconstruction. It enables extrapolation beyond scans via intersection-aware fitting. HybridSDF \cite{vasu_hybridsdf_2022} fuses analytic primitives with neural implicits using CSG-SDF operators. The method preserves sharp mechanical edges during generative modeling. BR-DF  \cite{zhang_b-rep_2025} encodes B-rep topology via per-face unsigned distance fields plus SDFs. Diffusion followed by Marching Cubes produces valid B-reps from the hybrid latent space.

    \subsection{AI Across Engineering Design}
    \label{ai-engdes}
    AI methods now appear in almost every stage of engineering design, but the types of models and their roles change as you move from problem definition to detailed design and communication.  Engineering design is often described in five stages: problem definition, conceptual design, preliminary design, detailed design, and design communication \cite{khanolkar_mapping_2023, allison_special_2022, yuksel_review_2023, afifi_data-driven_2025}. Khanolkar et al. \cite{khanolkar_mapping_2023} report that AI‑based methods are most frequently applied in conceptual and preliminary stages, with comparatively fewer applications in problem definition, detailed design, and especially design communication. A second set of surveys \cite{yuksel_review_2023} shows a similar trajectory, with a shift over the last decade from rule-based AI (expert systems, fuzzy logic) to data-driven ML/DL, and an increase in interest in explainable AI (XAI) for design. The five stages of engineering design and the use of AI methods in those stages are defined as follows:
        \subsubsection{Problem Definition}
        Designers identify objectives, constraints, functions, and requirements from customer needs, reviews, and specifications, combining empathy‑driven insight with data review. AI models such as NLP pipelines and classical ML (SVM, Naive Bayes, LDA) are used for customer-review mining, sentiment analysis, topic discovery, and requirement extraction. Data representations include textual inputs from reviews, patents, and specs, often tokenized into embeddings or graphs for semantic processing.
        \subsubsection{Conceptual Design} 
        Requirements are converted into engineering specifications; creativity and intuition are applied to ideate and generate novel design concepts and early evaluation. In conceptual design, deep learning methods such as CNNs, VAEs, and GANs, along with broader machine learning approaches, are frequently used for idea inspiration, 2D/3D concept generation, and feature‑ or function‑based recommendations. Other methods such as agent‑based modeling and probabilistic models also appear in selected applications, asreported in recent surveys of AI in engineering design. \cite{khanolkar_mapping_2023, allison_special_2022, yuksel_review_2023, kumar_singh_deep_2024}. Different tasks done using these models are:
        \begin{itemize}[leftmargin=*, nosep]
            \item Idea inspiration via semantic or technology networks constructed from patents and web text.  
            \item Generating 2D and 3D concepts from images, sketches, or functional descriptions using generative models (GANs, VAEs)\cite{krahe_ai-based_2019, chen_deep_2021, hoq_data-driven_2023, lee_deep_2022, wollstadt_carhoods10k_2022}. 
            \item Predicting function from form and recommending features or components.
        \end{itemize}
        Data representations consist of text-based knowledge graphs (e.g., ConceptNet \cite{speer_conceptnet_2018}), 2D sketches, and latent vectors from design repositories.
        
        \subsubsection{Preliminary Design}
        Engineers model concepts in CAD software and estimate sizes and attributes through analysis. AI accelerates this stage by rapidly generating and evaluating 3D models, allowing designers to explore larger design spaces efficiently \cite{khanolkar_mapping_2023, yuksel_review_2023}. Most widely used AI models in this stage are GANs and CNNs, alongside machine learning techniques like  decision trees and probabilistic methods.  
        Tasks performed using these models include:
        \begin{itemize}[leftmargin=*, nosep]
            \item Generative modeling, where GANs are utilized to produce 2D/3D concepts (e.g., tire treads \cite{lee_deep_2022}, aircraft \cite{chen_deep_2021}) from low-dimensional latent vectors, enabling design engineers to screen numerous variants quickly.
            \item Performance evaluation, where CNNs and supervised data-driven methods are trained on simulations or images/models to predict metrics like energy or stress, drag coefficients \cite{indupally_developing_2025, hoq_data-driven_2023, li_design_2023, pfaff_learning_2021}, replacing CFD/FEA to enable rapid feasibility checks and reduce computation time for iterative analysis.
        \end{itemize}
         Data representations include multi-view images, voxels, point clouds, B-Rep meshes, SDFs, and INRs
        \subsubsection{Detailed Design \& Design Communication}
        Designs are refined using codes and handbooks, with details optimized, followed by documentation and prototypes finalized for stakeholders. AI applications are less prevalent here compared to earlier stages, reflecting a reliance on established standards, handbooks, and manual review \cite{khanolkar_mapping_2023}.
        
        While AI demonstrates significant promise across conceptual and preliminary design stages, established geometric representations such as B-Rep, meshes, voxels, and point clouds, though effective for training ML models, they generally do not fully encode CAD semantics such as feature intent, parametric relations, and manufacturing constraints. A need thus arises for a representation that bridges native CAD designs to AI/ML tasks, enabling engineers to extract interpretable features, perform scalable learning (e.g., evaluation or surrogate modeling), and iteratively edit meaningful parameters that influence performance and manufacturability, all while remaining anchored in conceptual and preliminary phases. To address this gap, Section \ref{methodology-afg} defines a methodology for generating Attributed Feature Graphs (AFGs) from CAD models, demonstrated via an industrial case study of an automotive hood frame to illustrate the graph extraction process. Section \ref{case-study} then presents a case study applying this representation for surrogate modeling in preliminary design evaluation, training graph neural networks to predict structural performance and accelerate iterative analysis.


\begin{figure}
\centering\includegraphics[width=\linewidth, alt = {Outer Skin and Hood Frame}]{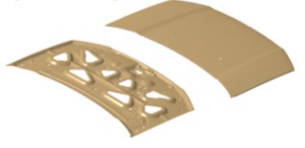}
\caption{Outer Skin and Hood Frame}\label{fig:outer_skin_frame}
\end{figure}
 

\section{Methodology: Attributed Feature Graphs}
\label{methodology-afg}
Feature‑based CAD modeling constructs complex geometry through parametric sequences of meaningful engineering operations (e.g., extrusions for ribs, offsets for depressions, cuts for pockets), organized as hierarchical feature trees that preserve shape and design intent. These trees encode manufacturing intent through their tree‑like topology, where root‑level primary features precede the children/secondary features.
While geometric representations (B‑Reps, voxels, meshes) excel at encoding final topology, they discard the feature semantics, limiting downstream learning, editing, and automation. Graph‑based representations offer a natural abstraction to capture these relationships, enabling learning‑based models to reason over topology, feature connectivity, and local geometry in a unified form. The goal is to define a feature‑based, CAD‑compatible representation that preserves design and manufacturing intent.

An Attributed Feature Graph (AFG), a CAD-native graph representation, is a graph representation of a design that divides the model based on the features and then encodes the geometric relationships between those features in a design context. An AFG represents feature-based CAD designs as a directed attributed graph \(\ G=\left(V,E,\operatorname{X}_V\right)\)\,  where nodes \(\ v_i \in V \)\ correspond to design features (depressions, ribs, pockets), node attributes \(\ \operatorname{x}_{v_i}\)\ encode explicit feature parameters (dimensions, local coordinates, size parameters), and edges \(\ e_{ij} \in E \)\ capture parent-child or geometric dependencies. AFGs mirror CAD feature trees but expresses them as structured, learnable representations. Unlike B-Reps \cite{jayaraman_uv-net_2021}, AFGs encode semantic features (slots, ribs, protrusions, etc.) rather than low-level topology (faces/edges). Unlike flat parameter vectors, AFGs preserve hierarchy and construction order, enabling bidirectional CAD to AFG translation. 
\par While the AFG definition above is independent of any product, its usefulness becomes clear when applied to a realistic feature-based design. This work uses automotive hood inner frames (Figure \ref{fig:outer_skin_frame}) as a case study to demonstrate how an AFG can be instantiated from an existing CAD model and then used in downstream learning tasks. Hood frames are especially well suited for this purpose because they are built as a combination of stamped features, expose a rich feature tree in modern CAD systems, and must satisfy stringent structural and manufacturing constraints. While the AFG construction pattern illustrated for hood frames is not specific to this application, it assumes that the CAD model contains a structured feature tree and explicit feature parameters. By relying on the same feature‑tree structure and CAD‑generation templates, any AFG that satisfies the feature‑compatibility constraints can be translated back into an editable/parametric CAD model.

\subsection{AFGs using Automotive Hood Frames}
Automotive hood inner frames exemplify complex feature-based design: thin-walled stamped panels integrating dozens of interdependent features (primary/secondary ribs, nested pockets, multi-level depressions) to balance stiffness, mass, and packaging within fixed outer skins. These designs are hierarchically structured, secondary ribs embed within primary depressions, pockets drive from surrounding flanges, making them ideal for demonstrating AFG construction as shown in Figure \ref{fig:afg_structure}. This instantiation validates AFG generality i.e. parse any feature-based CAD tree to extract parameters to infer hierarchy to store as learnable graph. CAD regeneration follows deterministically using feature templates organized as sketch–project–extrude–blend sequences, adapted from standard parametric‑CAD procedures. This ensures that optimized AFGs can be mapped back to parametric CAD models.

\subsubsection{CAD-AFG Conversion Framework}
\label{cad-afg-conversion}
In practice, the AFG is generated from a feature-based CAD model in three steps:
\begin{enumerate}
    \item Feature identification: depressions, pockets, and ribs are identified and classified from the CAD feature tree and surface geometry, using the feature classes and subclasses [Section \ref{features-hood}].
    \item Node construction: for each feature, its explicit record (class, local coordinate system, boundary descriptors, size parameters, activation state) is read from CAD and converted into the node attribute vector [Section \ref{def-nodes}].
    \item Edge construction: for each feature, its parent is determined from its support references (e.g., base surface, parent depression), and a corresponding directed edge is added [Section \ref{def-edges}].
\end{enumerate}

\begin{figure}[H]
\centering
\includegraphics[width=\linewidth, alt={Features Defined on a Hood Frame}]{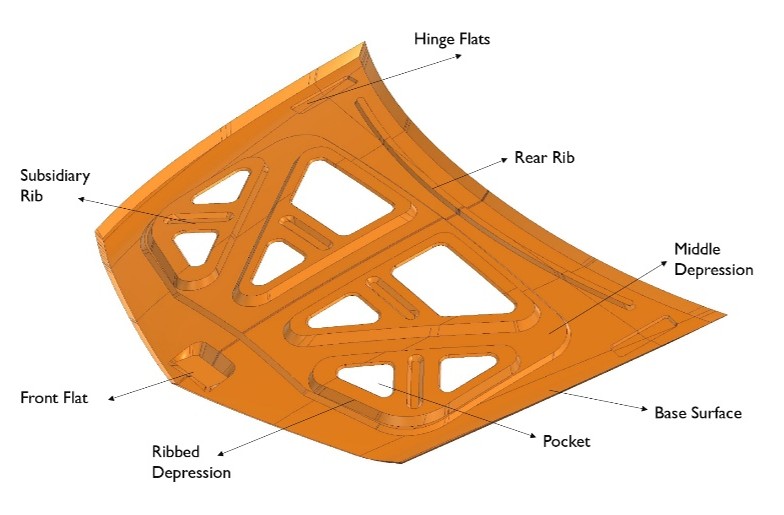}
\caption{Features Defined on a Hood Frame}\label{fig:features_hood_frame}
\end{figure}


This process yields a compact, semantically consistent graph representation in which each node corresponds to a mechanically meaningful feature, and each edge reflects a verifiable CAD support or dependency relationship. Because each node corresponds to a specific CAD feature, learned node and graph embeddings can be mapped back to individual features and their hierarchical contexts.

\subsubsection{Features on a Hood Frame} \label{features-hood} 
\begin{figure}
\centering\includegraphics[width=\linewidth, alt = {Feature Classification and Taxonomy}]{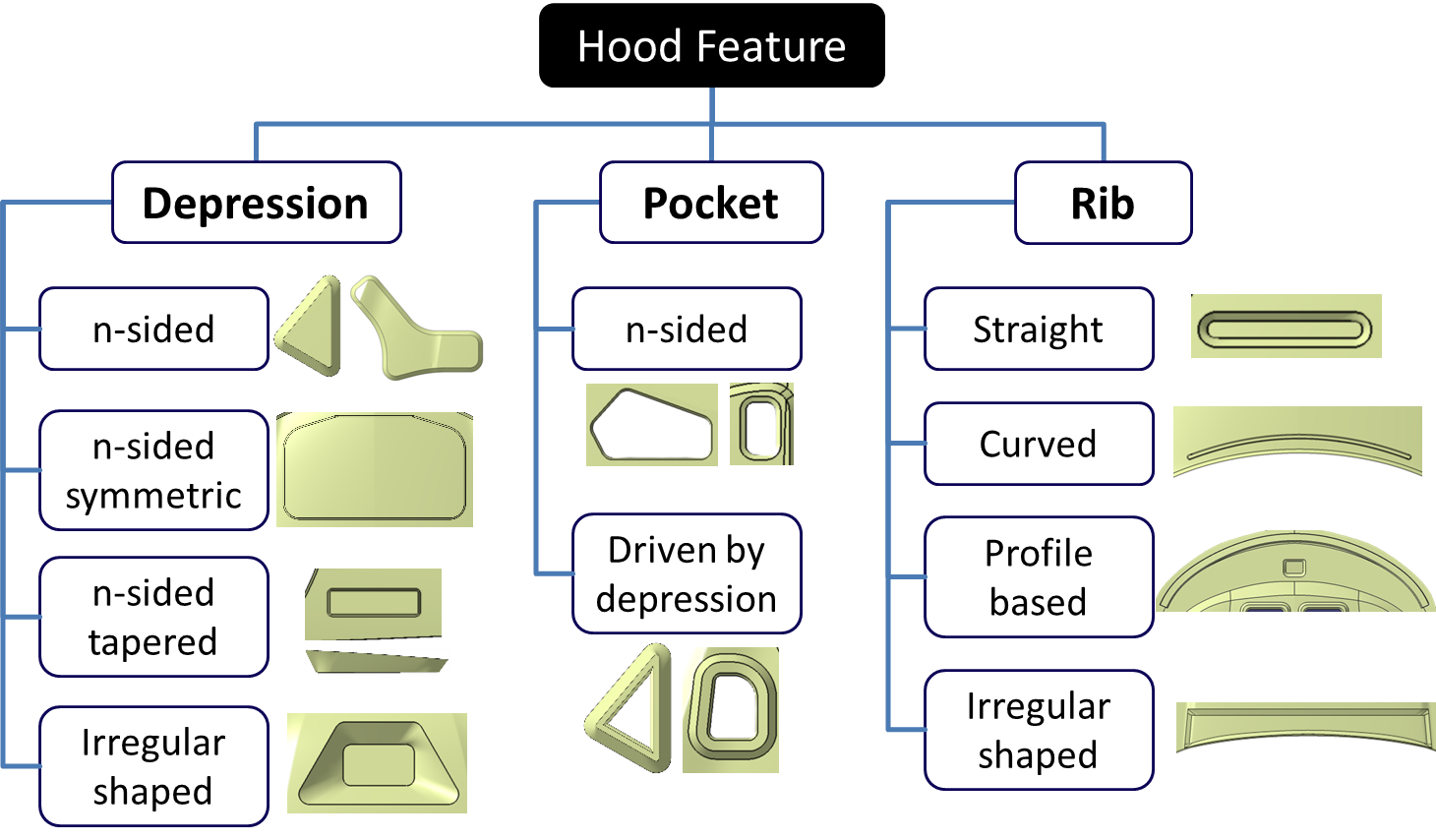}
\caption{Feature Classification and Taxonomy}\label{fig:feature_classification}
\end{figure}

To construct AFGs for hood frames, a feature ontology is defined - the canonical stamped features observed across production designs. Three primary classes  (Figure \ref{fig:feature_classification}) emerge from extensive analysis of reference hoods from an industry standard dataset, CarHoods10K \cite{wollstadt_carhoods10k_2022,noauthor_dryad_nodate}, a) Depressions: material pushed inward, including hinge/lock flats as special cases. b) Pockets: removed regions bounded by a flange lip (often used for mass reduction and local stiffness tuning). c) Ribs: raised stiffeners (straight or curved) that define dominant load paths and increase bending stiffness.
\par These classes correspond to the dominant manufacturable feature families used in hood inner frames and sufficiently represent the primary and secondary stiffening patterns. 
Feature levels in the AFG are hierarchically defined relative to parent surfaces, reflecting stamped manufacturing sequence (base → primary features → secondary features). Each level inherits geometric meaning from its parent, ensuring dependencies (e.g., pocket position tied to depression) are explicitly encoded. 

The following components are typical of the hood‑frame designs in the dataset being used and represent the core set of manufacturable features (as shown in Figure \ref{fig:features_hood_frame}) used to instantiate AFGs. The ontology can be extended if additional feature types appear in other designs.
\begin{itemize}[leftmargin=*, nosep]
    \item \textit{Base Surface}: It is the main surface supporting the hood frame’s features, typically derived from the hood skin's outer surface. This is considered the zero-level or base-level node in the graph structure.
    \item \textit{Front Flats}: They are secondary features that allocate space for a hood lock.
    \item \textit{Hinge Flats}: They are secondary features that allocate space for side hinges.
    \item \textit{Front Rib}: This is a secondary rib located on the front side of the hood frame.
    \item \textit{Rear Rib}: This is a secondary rib located on the back side of the hood frame.
    \item \textit{Middle Depression}: This is a first-level node whose supporting surface is always the base surface. It is always symmetric with respect to the hood frame center plane.
    \item \textit{Ribbed Depression}: This is a first-level or second-level node whose supporting surface is either the base surface or a middle depression. This is a compound feature, whose child nodes are pockets and subsidiary ribs.
    \item \textit{Pocket}: This is a third-level node whose supporting surface is a Ribbed depression.
    \item \textit{Subsidiary Rib}: This is a third-level node whose supporting surface is a Ribbed depression.
\end{itemize}



\subsubsection{Construction of Nodes \& Node Attributes} \label{def-nodes}

For hood frames, each node in the AFG represents a single stamped feature drawn from three primary classes: depressions, pockets, and ribs. Graph nodes can be added or removed as needed to compose the hood frame graph‑based representation, within the scope of the feature ontology defined earlier. A base surface node is used as the root. For every non‑root feature, a fixed‑length node attribute vector is constructed by collecting its explicit parameters from the feature record:
\begin{itemize}[leftmargin=*, nosep]
        \item \textit{Identity and type}: One‑hot or categorical encoding of feature class (depression, pocket, rib) and subclass (e.g., n‑sided depression, driven pocket, straight/curved rib).
        \item \textit{Placement}: Origin coordinates and local axes expressed in the parent’s local coordinate system, e.g., $(x_{orig}, y_{orig})$, which ensures that the feature can be reconstructed relative to its support region.
        \item \textit{Boundary geometry}: A reduced description of the feature footprint, such as normalized keypoint coordinates and corner radii for n‑sided polygons, or control‑point descriptors for curved ribs.
        \item \textit{Size parameters}: Dimensions that control structural behavior, such as rib length and width, depression flange width and level, and pocket flange width or offset.
        \item \textit{Activation flag}: A binary indicator for whether the feature is active in the current design, which allows discrete shape changes (feature insertion/suppression) to be represented without changing graph size.
\end{itemize}

\begin{figure}
\centering
\includegraphics[width=\linewidth, alt={Explicit Feature Record for a Ribbed Depression on a Hood Frame}]{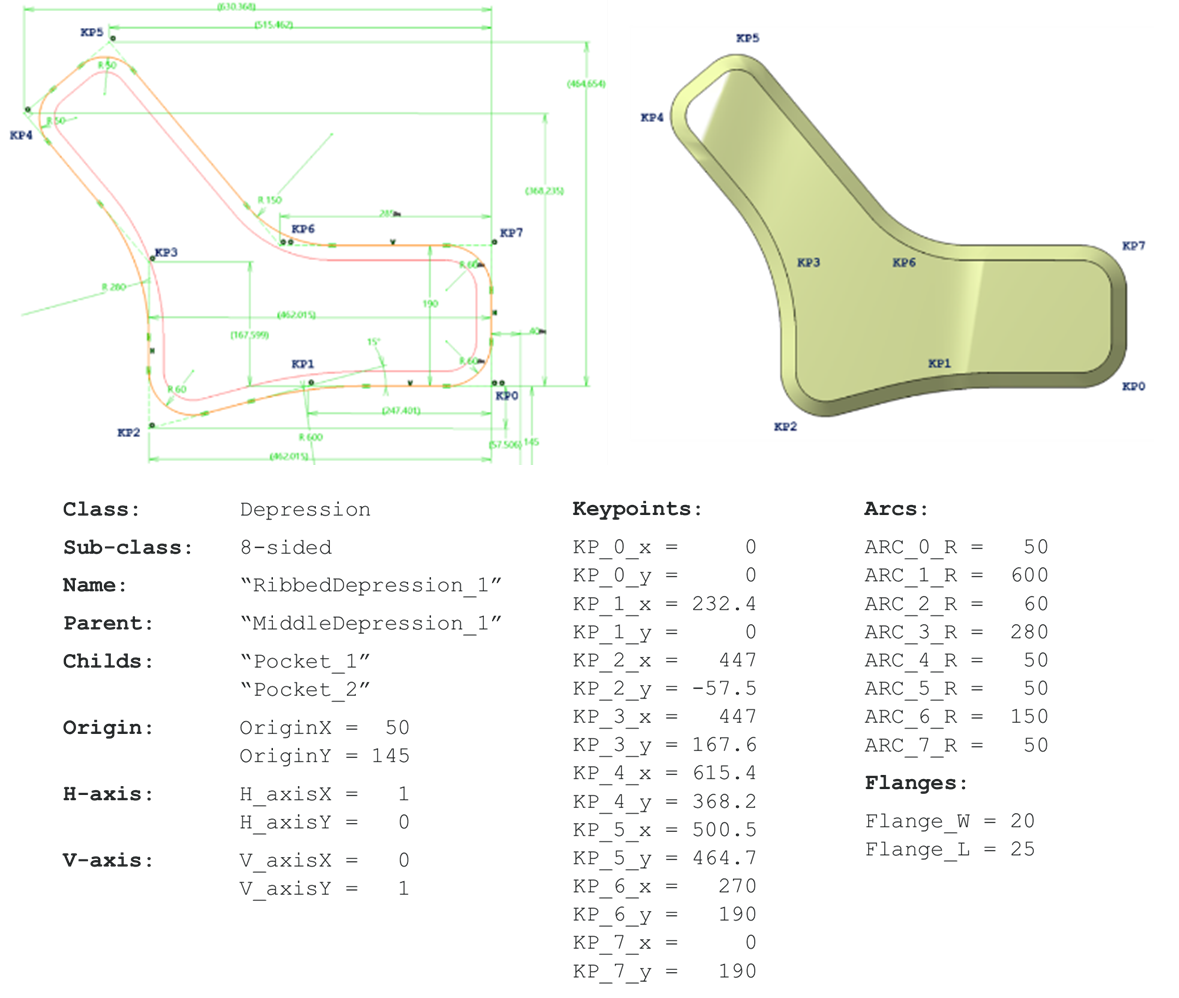}
\caption{Explicit Feature Record for a Ribbed Depression on a Hood Frame}\label{fig:explicit_record}
\end{figure}

All these quantities are read directly from the explicit feature representation. Feature class/subclass, local coordinate system, keypoint lists, and size parameters are extracted from the CAD model and mapped into a consistent attribute schema. Continuous parameters are normalized (e.g., by hood length/width), and categorical fields are embedded, yielding a numerical node feature vector. An example explicit feature record for a ribbed depression is shown in Fig.~\ref{fig:explicit_record}.

%

\subsubsection{Construction of Edges} \label{def-edges}

The AFG uses a hierarchy‑first adjacency convention that mirrors the feature tree stored in CAD. The base surface is the root node. First‑level children correspond to primary or functionally significant features such as depressions, hinge and latch regions, and dominant primary ribs. Second‑level children correspond to features defined on those primary regions, such as pockets embedded within a depression or subsidiary ribs defined relative to a ribbed region; additional levels are added as needed.

\begin{figure}
\centering\includegraphics[width=\linewidth, alt = {AFG Structure for a Hood Frame}]{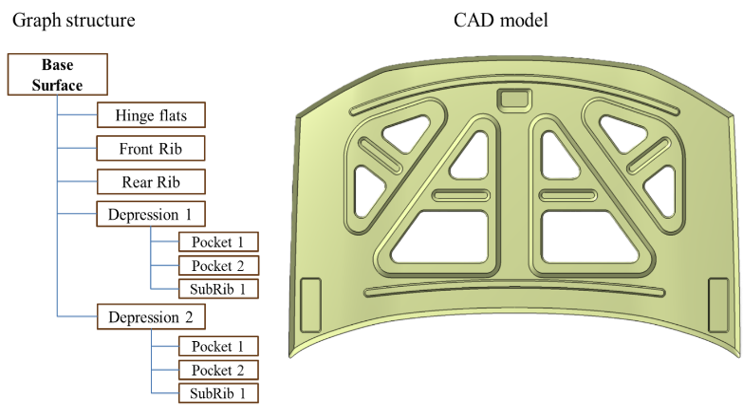}
\caption{AFG Structure for a Hood Frame; Nodes correspond to feature classes (depressions, pockets, ribs); edges encode parent‑child relationships.}\label{fig:afg_structure}
\end{figure}

Formally, the edge set $E$ is constructed as parent‑child edges, i.e., for every feature, a directed edge from its parent node to the feature node is added. The parent is determined from the CAD feature support (e.g., a pocket’s parent is the depression it is embedded in; a secondary rib’s parent is the parent feature, i.e., a primary rib or depression). At minimum, each edge stores the parent and child identifiers, which, together with the node attributes, are sufficient to reconstruct the hierarchy and provide a natural topological ordering for deterministic CAD regeneration (parents always precede children). From this edge list, an adjacency matrix $A$ is derived: $A_{ij}=1$ if there is an edge from node $i$ to node $j$, and $A_{ij}=0$ otherwise. Figure~\ref{fig:afg_structure} shows the generic hood frame feature‑tree showing the base surface, primary and secondary features.

With hood features defined and hierarchically organized, the following section describes the CAD‑AFG mappings that enable the full pipeline.


\begin{algorithm}
\caption{Forward Translator: AFG $\to$ CAD}
\label{alg:forward}
\begin{algorithmic}[1]
\REQUIRE AFG dataset ID $P$-$S$-$v$, base surface file $S$
\ENSURE Regenerated CAD hood‑frame part
\STATE Load base surface $S$, read AFG tables for $P$-$S$-$v$
\STATE Instantiate root node (base surface)
\FOR {each active feature node $n$ in AFG tables}
    \STATE Connect parent‑child edges using node IDs
\ENDFOR
\STATE \textbf{Traverse AFG in topological (parent‑first) order:}
\FOR {each node $n$ in traversal order}
    \STATE Create feature container, set local CSYS from origin relative to parent
    \STATE Select implicit template by $(class_n, subclass_n)$
    \STATE Execute template: sketch $\to$ project $\to$ blend/fill $\to$ join $\to$ split supports
    \STATE Update base/flange supports for child features
\ENDFOR
\STATE Finalize CAD part (naming, consolidation, export)
\end{algorithmic}
\end{algorithm}
\subsection{Bidirectional CAD‑AFG Mappings} 
The Attributed Feature Graph (AFG) enables a practical, deterministic mapping between feature‑based graph representations and editable CAD geometry, distinguishing it from lossy geometric representations (voxels, point clouds, B‑reps) that discard engineering intent and parametric structure.

\par \emph{Forward} mapping (AFG $\to$ CAD) regenerates feature‑based models via CAD‑neutral templates, as shown in Algorithm~\ref{alg:forward}; the regeneration workflow and its CAD-editability implications are demonstrated in\cite{alawadhi2026comparative}

\emph{Inverse} mapping (CAD $\to$ AFG) extracts AFGs via naming conventions and hierarchy parsing. This preserves semantic geometry and enables scalable, editable CAD representations for design workflows, as shown in Algorithm~\ref{alg:inverse}.

\begin{algorithm}[H]
\caption{Inverse Translator: CAD $\to$ AFG (Dataset Generation)}
\label{alg:inverse}
\begin{algorithmic}[1]
\REQUIRE CAD hood‑frame part, feature template rules
\ENSURE Stored AFG record (node/edge tables)
\STATE Open CAD part, identify base surface and global CSYS
\STATE Scan CAD tree/naming/geometry for candidate features
\FOR {each candidate feature $f$}
    \STATE Classify: assign $class_f$, $subclass_f$ via templates
    \STATE Extract explicit parameters: KP/CR lists, flange width/level, offsets, scales, active flag
    \STATE Compute local CSYS (origin) relative to parent support
\ENDFOR
\STATE Reconstruct edges: match parent‑child via template rules
\STATE Write AFG tables: nodes (parameters), edges (relations)
\end{algorithmic}
\end{algorithm}

\section{Case Study: Data‑Driven Performance Modeling using AFGs} \label{case-study}
Having established feature-based representations as the semantic bridge between CAD design intent and ML \ref{case-study}, their efficacy is demonstrated using an engineering case study. The CarHoods10K dataset \cite{wollstadt_carhoods10k_2022} provides parametrically varied hood frames with ground-truth FEA labels (maximum stress, mass, directional deflection), enabling direct comparison of surrogate models built on AFGs versus geometric baseline representations. This section details the model architecture of the evaluation engine, training setup, and results, serving as a concrete demonstration of how AFG‑based surrogates can support rapid, interpretable, CAD‑aware design iteration.
\begin{figure}
\centering\includegraphics[width=\linewidth, alt = {Evaluation Engine Architecture}]{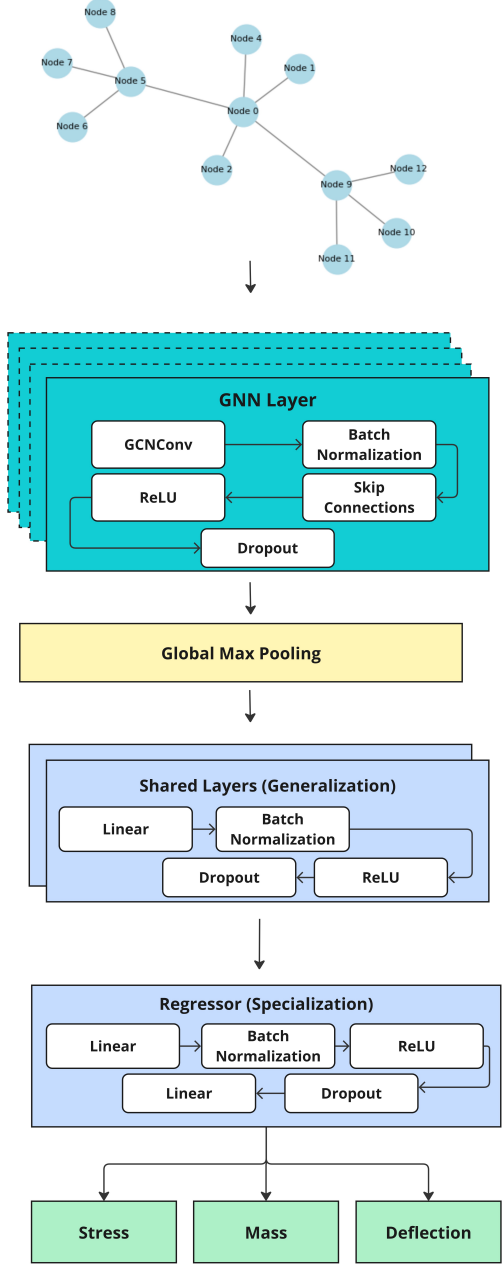}
\caption{Overview of the proposed multi-task graph neural network architecture. The input design is represented as a graph and processed by stacked GNN layers, followed by global max pooling, shared fully connected layers, and task-specific regression heads for predicting stress, mass, and deflection.}\label{fig:6}
\end{figure}
 
\subsection{Evaluation Engine Architecture}
Each structure is represented as an directed graph
\(G = (V,E),\ \)where nodes~\(v \in V\)~correspond to features on the
hood frame and edges~\(e = (u,v) \in E\)~represent geometric
relationships between features (e.g., connected features, location from
parent feature).

Through message passing, each node iteratively updates its feature
vector using information received from its neighbors. After several
layers, a node's representation no longer captures only its own
attributes but also aggregated information from nodes multiple hops away
in the graph. As a result, the learned embeddings are context-aware:
they encode both local properties and the broader structural
neighborhood, which is crucial for accurately predicting global
responses (e.g., stress, mass, or deflection) from the graph.

In this work, message passing is implemented using stacked GCNConv
\cite{kipf_semi-supervised_2017} layers, which realize a specific instance of the above
framework. For a single layer, the update rule can be written as
\begin{equation}
  H^{(k)} = \sigma\left( {\widetilde{D}}^{- \frac{1}{2}}\widetilde{A}{\widetilde{D}}^{- \frac{1}{2}}H^{\left( k-1 \right)}W^{(k)} \right)
\end{equation}

Where, \(H^{(k)} \in \mathbb{R}^{\mid V \mid \times d_{k}}\) stacks node
features~\(h_{v}^{(k)}\), \(\widetilde{A} = A + I\) is the adjacency
matrix with self-loops, \(\widetilde{D}\) is the corresponding degree
matrix, \(W^{(k)}\) is a learnable weight matrix, and
\(\sigma( \cdot )\) is a non-linear activation. This operation can be
viewed as a learned, degree-normalized average of each node's neighbors
followed by a linear transformation and nonlinearity, enabling the
network to learn how local neighborhoods (feature interactions)
contribute to structural performance. The backbone of the model consists
of~\(L\)~identical GNN blocks. Each block applies a GCNConv layer
followed by batch normalization, a ReLU activation, and dropout.
Residual (skip) connections are added around the block to stabilize
optimization and mitigate over-smoothing in deeper stacks by allowing
gradients and information to bypass individual layers. Formally, for
block~\(k\),
\begin{equation}
    {{\widehat{H}}^{(k)} = Dropout\left( ReLU\left( BN\left( GCNConv\left( H^{\left( k-1 \right)} \right) \right) \right) \right)}
\end{equation}
\begin{equation}
    {H^{(k)} = H^{\left( k-1 \right)} + {\widehat{H}}^{(k)}}
\end{equation}

After the final GNN block, node embeddings are pooled into a fixed-size
graph-level representation using global max pooling, yielding a graph‑level embedding
\(z_{j} = \underset{v \in V}{max} \space h_{v,j}^{(L)},\) where
\(z \in \mathbb{R}^{d_{L}}\) is the latent vector summarizing the most
important response-relevant features across all nodes and channels. This
pooling operation enables architecture to handle graphs of varying sizes
while emphasizing the strongest activating structural patterns. The
pooled representation is passed through a shared fully connected
subnetwork that captures features common to all prediction tasks. 


This shared head consists of a linear layer, batch normalization, ReLU
activation, and dropout, yielding a task-agnostic latent vector~\(g\).
On top of this shared representation separate regression heads for
stress, mass, and deflection are added, implemented as small multilayer
perceptron (Regressor) as shown in the figure \ref{fig:6}. The shared trunk
promotes parameter efficiency and inductive transfer between related
structural response predictions, while the task-specific heads allow
specialization to each output quantity. The following algorithm explains
the forward propagation for this model:

Algorithm ~\ref{alg:architecture} details the forward pass, implemented using PyTorch Geometric GCNConv layers with global max-pooling for graph-level aggregation.
\begin{algorithm}
\caption{Evaluation Engine Forward Pass}

\label{alg:architecture}
\begin{algorithmic}[1]
\REQUIRE Graphs $G_i$, node features $\mathbf{X} \in \mathbb{R}^{N \times F}$, edge index $\mathbf{E}$, batch vector $\mathbf{b} \in \mathbb{R}^N$, depth $K$, GCN weights $\mathbf{W}_k$, skip projections $\mathbf{S}_k$ (optional), BatchNorm layers, dropout $p$, flag
\ENSURE Predictions $\hat{\mathbf{y}}_g$ for graphs and targets

\STATE Initialize node embeddings: $\mathbf{H}^{(0)} = \mathbf{X}$

\FOR{$k = 1$ to $K$}
    \STATE Store identity: $\mathbf{I} = \mathbf{H}^{(k-1)}$
    \STATE Graph convolution + normalization:
    \[
    \mathbf{H}^{(k)} = \text{BN}_k \left( \hat{\mathbf{A}} \mathbf{H}^{(k-1)} \mathbf{W}_k^{(1)} \right)
    \]
    \STATE Residual connection (if applicable): $\mathbf{H}^{(k)} \leftarrow \mathbf{H}^{(k)} + \mathbf{S}_k(\mathbf{I})$
    \STATE Non-linearity: $\mathbf{H}^{(k)} \leftarrow \text{ReLU}(\mathbf{H}^{(k)})$
    \STATE Dropout (if): $\mathbf{H}^{(k)} \leftarrow \text{Dropout}(\mathbf{H}^{(k)}, p)$
\ENDFOR

\STATE Global max pooling: $\mathbf{h}_g = \max_{i \in G} \mathbf{H}^{(K)}_i$
\STATE Shared MLP: $\mathbf{h}_g \leftarrow \text{MLP}_{\text{shared}}(\mathbf{h}_g)$ \quad (2-layer, BN, ReLU, Dropout)
\STATE Task heads: $\hat{\mathbf{y}}_g = \text{MLP}_{\text{head}}(\mathbf{h}_g)$ \quad (stress/mass/deflection)
\RETURN $\hat{\mathbf{y}}_g$ as regression output
\end{algorithmic}
\end{algorithm}

\subsection{Training \& Loss Function}
\begin{figure}[H]
    
    \centering
    \begin{tabular}{c c}
    
    \begin{minipage}[t]{0.5\columnwidth}
    \centering
    \includegraphics[width=\linewidth]{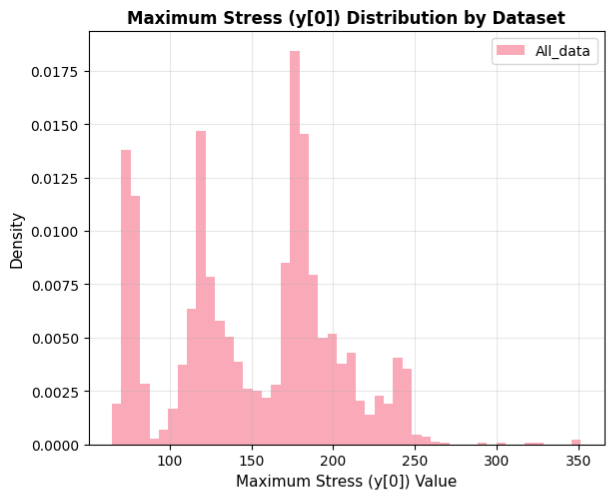}
    
    \vspace{0.5ex}
    \textbf{(a)} Data Distributions (Stress)
    \end{minipage}
    &
    \begin{minipage}[t]{0.5\columnwidth}
    \centering
    \includegraphics[width=\linewidth]{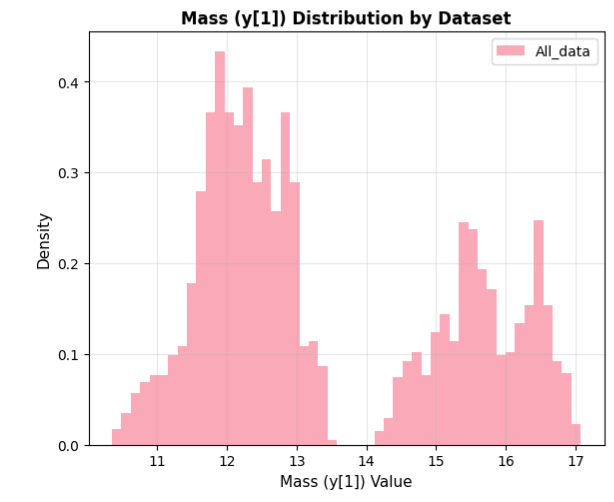}
    
    \vspace{0.5ex}
    \textbf{(b)} Data Distributions (Mass)
    \end{minipage}
    \\
    \\
    
    \multicolumn{2}{c}{
    \begin{minipage}[t]{1\columnwidth}
    \centering
    \includegraphics[width=0.5\linewidth]{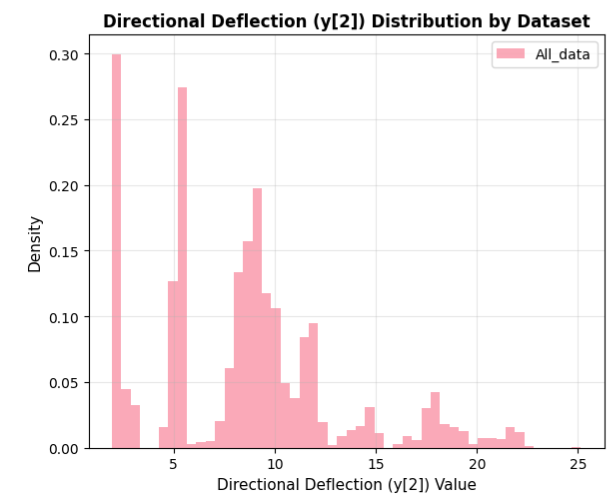}
    
    \vspace{0.5ex}
    \textbf{(c)} Data Distributions (Deflection)
    \end{minipage}
    }
    \\
    
    \end{tabular}
    \caption{Data Distributions}
    \label{fig:data_distributions}
    \end{figure}

The CarHoods10K dataset comprises over 10,000 designs generated from 10 base skins $\times$ 10 feature patterns $\times \sim 100$ size variations each. For training, only the 6 industrially representative skins and their corresponding 6 feature patterns were retained to ensure manufacturability, yielding $6 \times 6 \times \sim 100 = \sim 3{,}600$ designs, of which 3,005 had valid performance metrics. The dataset was split into 2,103 training, 601 validation, and 301 test samples (approximately 70/20/10\% of the retained designs), with the test set chosen to be geometrically distinct from the training set. Figure \ref{fig:data_distributions} shows that the target responses span a broad range, with clear spread, skewness, and outliers in maximum stress, mass, and directional deflection. This indicates that the surrogate model is trained on diverse engineering outcomes rather than a narrow response regime.

A single multi-task surrogate model, referred to as the \emph{Evaluation Engine}, maps each AFG to a three-dimensional output vector:
\[
\bigl[y_{\text{stress}},\, y_{\text{mass}},\, y_{\text{deflection}}\bigr].
\]
The model uses a GCNConv-based backbone with a shared MLP and three task-specific regression heads. It contains 270{,}339 trainable parameters and handles variable-size graphs through global max pooling. The model is trained on mini-batches of graphs sampled from the dataset, with batch normalization and early stopping used to stabilize optimization.

Training computes per‑target MSE against ground truth, then takes a weighted sum before the backward pass:
\begin{equation}
\mathcal{L} = \sum_{i=1}^{3} w_i \cdot \mathrm{MSE}(y_{\text{metric}_i}, \hat{y}_{\text{metric}_i})
\end{equation}
where metric $\in$ [stress, mass, deflection], $y_{\text{metric}_i}$ is the predicted value, $\hat{y}_{\text{metric}_i}$ is the actual value, and $\mathrm{MSE}$ is the mean squared error. The weights $w_i$ are chosen to balance the influence of each task in the total loss.

The three targets differ in engineering importance and have different numerical scales; to separate these concerns, all outputs are first normalized using a RobustScaler (median and interquartile range) computed on the training‑set outputs only, to avoid data leakage. After normalization, the per‑task validation losses were measured under an unweighted objective, and the final weights were set to the (unnormalized) inverse of these validation losses. There was no further manual scaling or adjustment of the weights after this step, so the loss weights are fully data‑driven and derived from the observed validation‑set performance:
\[
[w_{\text{stress}}, w_{\text{mass}}, w_{\text{deflection}}] = [2.5, 1.0, 4.5].
\]
This choice prevents any single task from dominating the loss while still reflecting the relative difficulty of the targets, with higher weight on deflection due to its larger raw validation‑set error.

The final model used a 4-layer GCN backbone with 256 hidden channels, dropout of 0.3, and skip connections enabled. Training was performed with Adam using a learning rate of 1e-3, weight decay of 2e-4, and batch size 32 for up to 1500 epochs. Early stopping with a patience of 45 epochs and a ReduceLROnPlateau scheduler were used to stabilize training and select the best checkpoint. Training was performed on an NVIDIA RTX 2000 Ada GPU (16 GB VRAM) with an Intel i7‑14700K CPU; typical runs required 250–300 epochs ($\approx$ 1–1.5 hours), with early stopping performed at the epoch achieving the lowest validation loss ($\approx$ 220 epochs).

\begin{figure*}[htbp]
    \centering
    \begin{subfigure}[b]{0.32\textwidth}
        \centering
        \includegraphics[width=\linewidth]{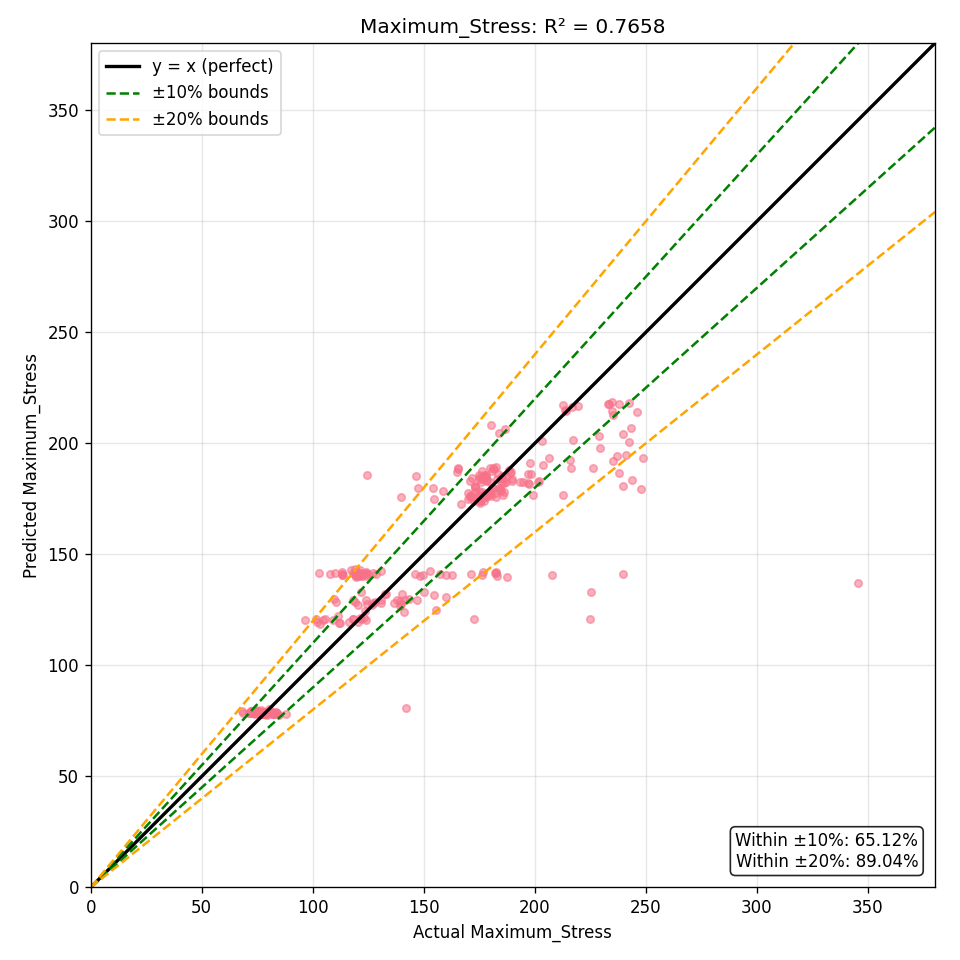}
        \caption{Maximum Stress}
    \end{subfigure}
    \hfill
    \begin{subfigure}[b]{0.32\textwidth}
        \centering
        \includegraphics[width=\linewidth]{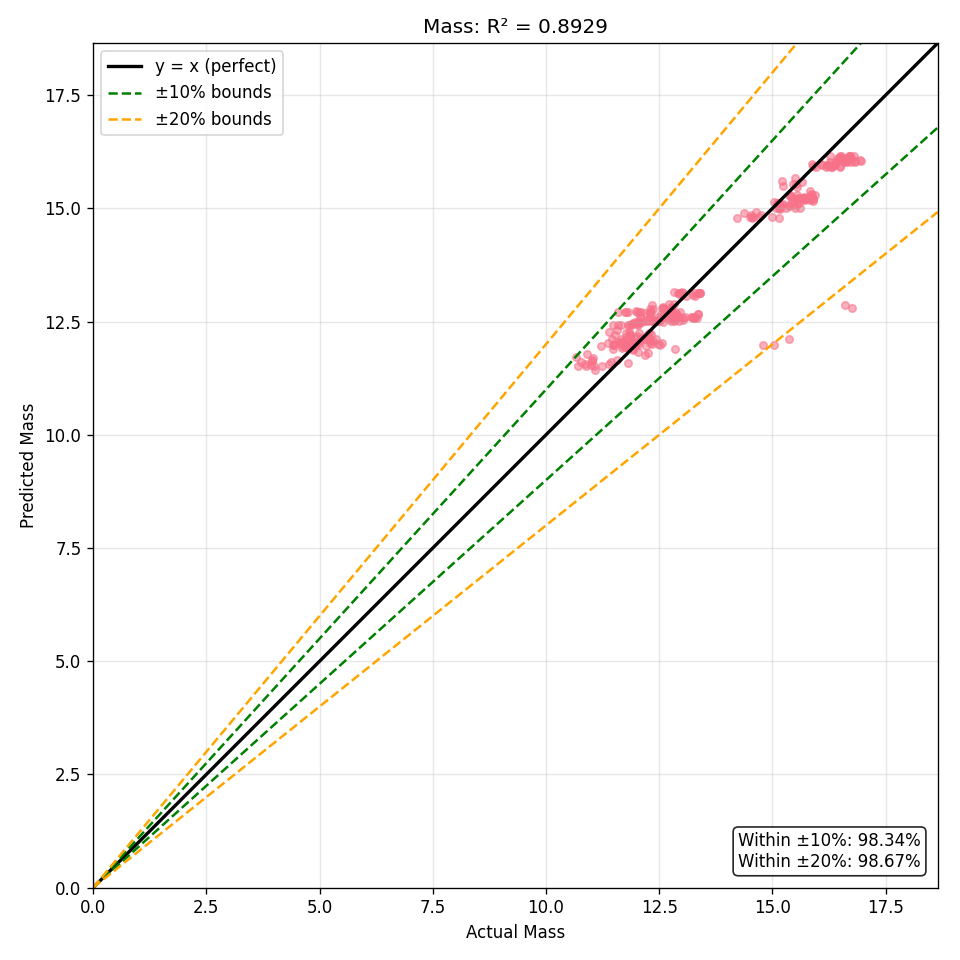}
        \caption{Mass}
    \end{subfigure}
    \hfill
    \begin{subfigure}[b]{0.32\textwidth}
        \centering
        \includegraphics[width=\linewidth]{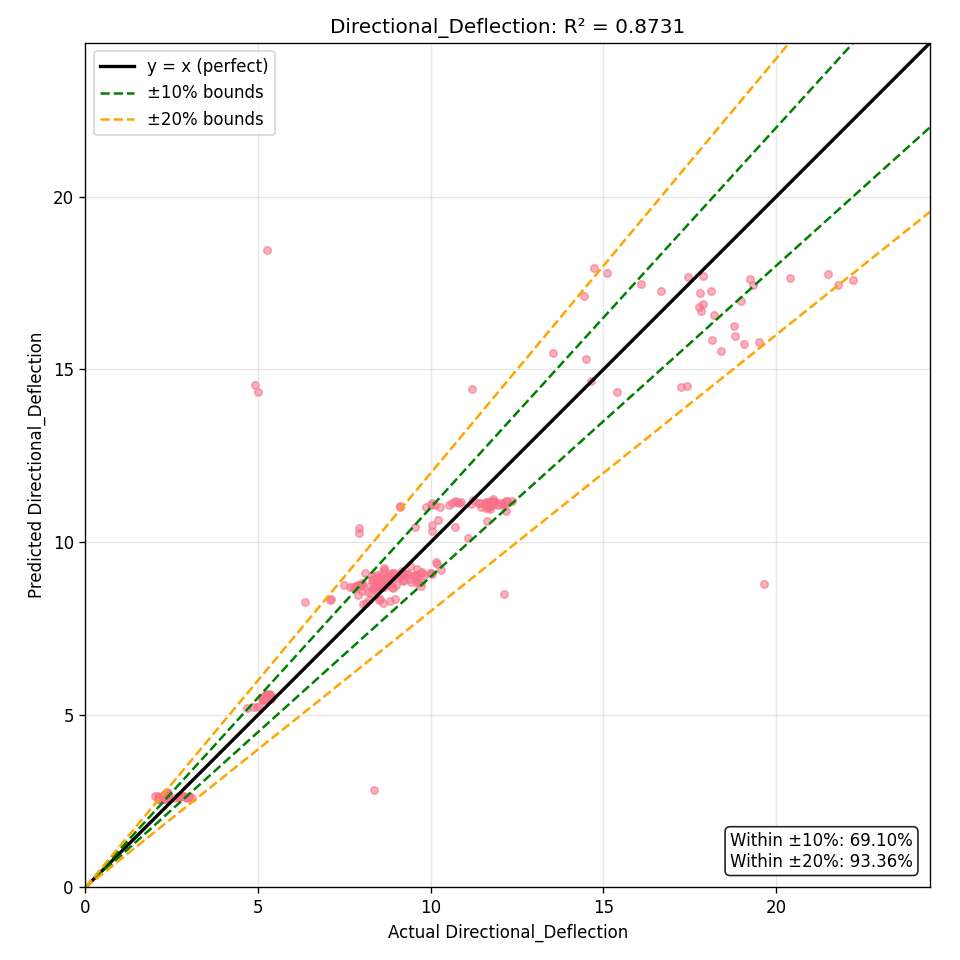}
        \caption{Directional Deflection}
    \end{subfigure}
    \caption{Test‑set $R^2$ distributions for maximum stress, mass, and directional deflection predictions by the Evaluation Engine.}
    \label{fig:r2_plots}
\end{figure*}
\subsection{Results}
Test metrics demonstrate robust predictions (Table~\ref{tab:train-metrics}): overall MAPE on normalized targets is approximately 9.4–9.6\%, with ${R}^2 = 0.84$ and 88\% of predictions within 20\% of ground truth. Mass predictions are particularly strong (${R}^2 = 0.89$, 98.67\% within 20\%), stress predictions are solid (${R}^2 = 0.77$, 89.04\% within 20\%), and deflection predictions are moderate but still suitable for engineering surrogates (${R}^2 = 0.87$, 93.36\% within 20\%).
Error bands in the plots (Figure \ref{fig:r2_plots} confirm consistent performance across the full operating range, with no systematic bias observed. Overall, 93.6\% of predictions fall within 20\% across all outputs, validating the model's generalization for engineering surrogate applications.

\begin{table}[htbp]
\centering
\caption{Training and Test Metrics}
\label{tab:train-metrics}
\resizebox{\linewidth}{!}{%
\begin{tabular}{lcccc}
\toprule
\textbf{Metric} & \textbf{Train $R^2$} & \textbf{Test $R^2$} & \textbf{Within 10\%} & \textbf{Within 20\%} \\
\midrule
vM Stress      & 0.79 & 0.76 & 65.12\% & 89.04\% \\
Mass        & 0.90 & 0.89 & 98.34\% & 98.67\% \\
Deflection  & 0.84 & 0.87 & 69.10\% & 93.36\% \\
\bottomrule
\end{tabular}}
\end{table}

\begin{table*}[t]
\centering
\small
\caption{Comparison of ML Models on the CarHoods10K Test Set}
\label{tab:comparison}
\begin{tabular}{p{1.4cm} p{2.6cm} p{2.2cm} ccc cc cc}
\toprule
\textbf{Model} & 
\textbf{Data} & 
\textbf{Architecture} &
\multicolumn{6}{c}{\textbf{Performance Metrics (Test Set)}} \\
\cmidrule(lr){4-9}
& & & 
\multicolumn{2}{c}{\textbf{vM Stress}} &
\multicolumn{2}{c}{\textbf{Mass}} &
\multicolumn{2}{c}{\textbf{Deflection}} \\
\cmidrule(lr){4-5} \cmidrule(lr){6-7} \cmidrule(lr){8-9}
& & &
MAPE & $R^2$ &
MAPE & $R^2$ &
MAPE & $R^2$ \\
\midrule
CNN \cite{kumar_study_2024} &
Images (Top + Side Views) &
ResNet50 &
8.20\% & -- &
8.10\% & -- &
15.20\% & -- \\
\midrule
MMML \cite{indupally_developing_2025} &
Images + Rib Depths + Cross Sections &
ResNet50 + MLP &
3.60\% & -- &
4.70\% & -- &
10.50\% & -- \\
\midrule
PC-VAE \cite{wollstadt_carhoods10k_2022} &
Point Clouds &
VAE &
7.96\% & 0.82 &
3.51\% & 0.94 &
8.25\% & 0.94 \\
\midrule
GNN (\textit{current}) &
Attributed Feature Graphs (AFGs) &
GCNConv + MLP &
9.18\% & 0.76 &
2.98\% & 0.89 &
10.58\% & 0.87 \\
\bottomrule
\end{tabular}
\end{table*}

\section{Discussion}\label{discussion}
\begin{figure*}[t]
    \centering
    \includegraphics[width=\textwidth, alt={Engineering Design Workflow using AFGs}]{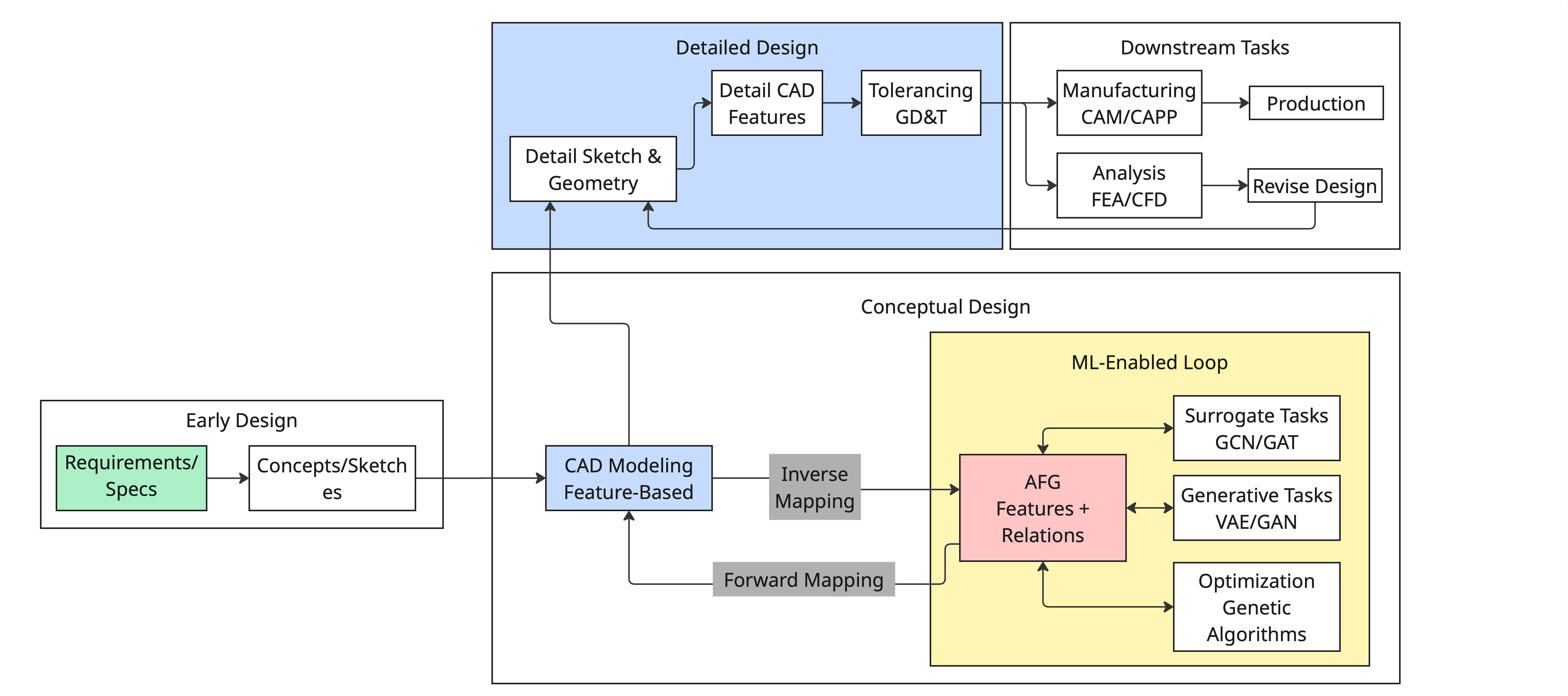}
    \caption{Engineering Design Workflow using AFGs}
    \label{fig:ed_workflow}
\end{figure*}
 
This work positions Attributed Feature Graphs (AFGs) as an alternative CAD-centric geometric representation for machine learning in engineering design, complementing existing image-, point-cloud-, and purely parametric approaches. Rather than focusing solely on predictive accuracy, this research argues that the utility of a design representation lies in how well it bridges learned behavior with editable design intent. Evaluated on the CarHoods10K dataset, different representations reveal distinct trade-offs in both performance and interpretability. Baseline results are reported as published in the corresponding studies; because the exact train/validation/test splits and filtered subsets are not always available, the comparison should be interpreted as a benchmark against reported prior results rather than a fully controlled re-evaluation under identical data partitions.

CNNs operating on 2D images attain moderate accuracy (Table~\ref{tab:comparison}, row CNN), but fully decouple geometry from its parametric origin, severing any direct link to editable CAD parameters. The multi-modal fusion approach (MMML) improves these metrics but relies on manually extracted geometric descriptors (rib depths, cross‑sections), which constrains scalability and generalizability across evolving design families. PC‑VAE on point clouds excels in mass prediction due to dense geometric fidelity but under performs on stress and deformation compared to some alternatives. In contrast, the proposed Attributed Feature Graphs (AFGs), processed by the \emph{Evaluation Engine} (GNN‑MLP), yield competitive performance across all three targets while encoding designs as sparse, semantically rich graphs rooted in CAD features such as extrusions, depressions, ribs, and cutouts. This structure makes AFGs a viable geometry representation for ML while organizing CAD designs in a feature-based graph form that is more directly aligned with engineering editability than pixel arrays, point clouds, or raw parameter vectors. While the results support the proposed framework, a full ablation study isolating hierarchy, feature semantics, and node attributes remains future work.

Beyond numeric accuracy, the distinctive value of AFGs lies in how they align with engineering design workflows. Because each graph node maps directly to an editable CAD feature, predictions can be interpreted in terms of actionable design primitives, enabling engineers to trace performance outcomes to specific design choices. This interpretability should be understood as model interpretability rather than direct physical causality: attribution methods such as SHAP or GNNExplainer can identify which CAD features most influenced a prediction, but those attributions do not by themselves establish causal physical effects. The CAD-editability and bidirectional CAD–AFG mapping of this workflow are demonstrated in \cite{alawadhi2026comparative}. For example, when the AFG-based prediction model identifies a rib feature as strongly affecting stress predictions, the designer can modify parameters such as rib thickness within the native CAD environment and obtain updated surrogate estimates within seconds. This supports a closed AI–CAD design loop, where learning models inform iterative design modifications while remaining grounded in the same CAD semantics used by human designers.

AFGs offer a CAD-aware representation for machine learning that preserves semantic structure, design intent, and editability, rather than reducing geometry to an unstructured or purely geometric signal. We do not claim that AFGs are universally superior to existing representations; rather, we show that they broaden the set of design-centric encodings available for engineering tasks such as performance prediction, feature-level sensitivity analysis, and design space exploration. In this sense, AFGs provide a complementary pathway for integrating machine learning with CAD-driven design, as illustrated in Figure \ref{fig:ed_workflow}.


\subsection{Limitations and Future Work} \label{limitations-future-work}

Although AFGs provide a promising bridge between CAD and data‑driven modeling, the current framework inherits several limitations common to geometry‑based ML in engineering design. Below is a summary of limitations of the representation and outlines future directions that extend beyond surrogate modeling toward broader design-centric ML use cases.

\subsubsection{Limitations}

\begin{itemize}[leftmargin=*, nosep]
  \item AFGs encode empirical patterns without explicit physical constraints, leaving physical consistency to downstream models or post-processing.

  \item Defining the initial AFG ontology (e.g., “rib” vs. “web”) requires domain expertise and manual effort before any ML model can be applied.

  \item Gradient‑based analyses over AFGs can be misleading under out-of-distribution designs, as learned gradients may not reflect true physical or design sensitivities in unseen regions.
  
  \item Representation quality is limited by data coverage; richer datasets spanning more design families and manufacturing constraints are needed.

  \item Requires standardized CAD naming and modeling workflows for dataset creation, because inconsistencies can disrupt graph dataset generation.

\end{itemize}

\subsubsection{Future Work}

\begin{itemize}[leftmargin=*, nosep]

  \item Explore generative and meta‑learning models built on AFGs to enable interpretable, feature‑level design space exploration, design generation, and cross‑product generalization.

  \item Investigate physics-aware models using AFG representations that couple empirical patterns with design or physical constraints.
  
  \item Integrate AFG-based workflows into interactive CAD environments for real-time, editable design feedback and exploration.
  
  \item Investigate ablation studies and representation variants to isolate the contributions of graph hierarchy, node attributes, and pooling strategies within AFG-based surrogate and generative models.


\end{itemize}
\section{Conclusion}

This work introduces Attributed Feature Graphs (AFGs) as a new, feature-based representation for CAD designs, using a surrogate modeling task on the CarHoods10K dataset as a case study to demonstrate their utility. By encoding CAD features such as extrusions, ribs, and cutouts as attributed nodes and edges, AFGs provide a sparse, interpretable structure that preserves design intent and supports direct mapping to editable parameters in CAD. Across stress, mass, and deformation prediction, AFGs yield competitive performance while enabling engineers to trace model outputs to concrete design primitives and iterate rapidly within native CAD environments.

More broadly, this work emphasizes that the way engineering design data is represented fundamentally shapes how machine learning models learn, generalize, and support decision‑making. While modern ML methods can capture complex patterns through diverse mathematical formulations, engineers must still anchor design choices in experience and knowledge that cannot be fully encoded as explicit rules. AFGs sit at the interface between these two worlds: by representing CAD designs as feature‑based graphs that mirror human design intent, they provide models with structured, interpretable inputs and make it easier for designers to trace performance outcomes to specific geometry features.

Feature‑based representations such as AFGs thus shift the emphasis from purely “black‑box” prediction toward interpretable, design‑centric representations. Because nodes and edges correspond directly to editable CAD features, predictions can be understood and edited within the same parametric environment used in practice. This tight coupling between representation, model, and design workflow enables engineers to leverage data‑driven methods while retaining control over the design rationale.

In summary, AFGs offer a complementary way to represent CAD geometry for ML in engineering design, not as a universal best representation, but as one that aligns geometry, semantics, and editability. How data is represented remains a central concern in learning‑based design; by choosing representations that reflect design structure and intent, models can be built in a way that are not only accurate but also interpretable and actionable for real engineering workflows.


\section*{Acknowledgments}
The authors would like to acknowledge the financial and technical support received from Honda Research Institute - US and Honda Development and Manufacturing Americas LLC.


\bibliographystyle{unsrtnat}
\bibliography{IDETC_26}

\end{document}